\documentstyle[12pt,epsf]{article}
\begin{document}
%\rightline{UWThPh-1998-51}
\begin{center}
{\Large \bf Nonperturbative interaction in $q \bar{q}$ bound states\\}
\bigskip
{\large N. Brambilla$^a$  and A. Vairo$^b$\\}
\date{}
\smallskip
{\it $^a$ Institut f\"ur Theoretische Physik, Boltzmanngasse 5, A-1090, Vienna, Austria\\}
\smallskip
{\it $^b$ Institut f\"ur Hochenergiephysik, \"Oster. Akademie der Wissenschaften\\
 Nikolsdorfergasse 18, A-1050 Vienna, Austria}
\end{center}

\begin{abstract}
We report about recent progress in the treatment of bound states in QCD.
\end{abstract}

\section{Introduction}
One of the most difficult and less understood problems in Quantum Field Theory is the treatment of 
bound state systems. 
 Even in QED where the interaction is simply given in perturbation theory, 
the practical  calculation of the bound state properties  is  tricky.
The  complication comes  from the mixing of the many characteristic scales of the bound state,
the mass of the   particles, the  momentum and the energy of the state.  
A way to handle this problem is provided by Non Relativistic QED (NRQED) which  
  is an effective theory equivalent to QED  and obtained from QED by integrating out the hard 
energy scale  $m$. The ultraviolet regime of QED  (at energy scale $m$) is perturbatively encoded
order by order  in the coupling constant $\alpha$  in the matching coefficients that appear
in front of the operators in the effective Lagrangian. Each term in the effective Lagrangian 
has a definite power counting  in $\alpha$ and then a disentangling of the various scales is 
achieved. A similar simplification is obtained using Non Relativistic QCD (NRQCD) in heavy quark
systems. These systems are characterized by a dynamical dimensionless parameter, the quark 
velocity $v$, which is small  and allows a classification of the energy scales of the 
problem in hard ($\sim m$), soft ($\sim mv$) and ultrasoft ($\sim mv^2$). This provides a 
power counting scheme.  Here, however,  a further and  conceptual difficulty  arises in connection
 with the nonperturbative nature of low-energy QCD.
The relation between $v$ and the QCD parameters is unknown but 
certainly $v$ includes both perturbative and nonperturbative effects.  Therefore,  the evaluation 
of the bound state properties is  ultimately done with a lattice simulation.
 In the case of bound systems with at least one light quark  simplifications of this type 
do not hold.   On one hand  no  small expansion parameter  exists, on the other hand  
the light quark mass is generated via chiral symmetry breaking. Hence, the interplay 
between confinement and chiral symmetry breaking has to be considered.\par
 In this talk we report about  recent progress in the treatment of bound states in QCD.
We show that   it is possible to obtain a
 model-independent and gauge-invariant result for the 
heavy quark interaction  at  order $v^4$ of the systematic expansion in $v$.
The interaction turns out to be simply given in terms of a generalized (distorted) Wilson loop.
The result is suitable for lattice evaluation as well as for analytic evaluation 
once a QCD vacuum model is considered.  We show that the results for the heavy quark dynamics are
substantially under control and are given in terms of two nonperturbative parameters $T_g$, the 
gluon correlation length,  
and $G_2$, the gluon condensate.
Adopting  the same framework  in order to study the heavy-light bound states in the non-recoil 
limit,   spontaneous chiral symmetry breaking and a confining chiral non-invariant 
interaction emerge quite naturally.  We discuss this last case with more details.

\section{The heavy quark interaction}

The heavy quark interaction can be obtained analytically  at the order $v^4$ of the   
systematic expansion of the interaction in $v$. Here, we report only the main steps of the derivation 
referring to \cite{nora,hugs} for further details. 
\begin{itemize}
\item{\it Step 1}. Set up the NRQCD Lagrangian.
The  NRQCD Lagrangian \cite{nrqcd} 
is obtained from the QCD Lagrangian via a Foldy-Wouthysen transformation.
At order $O(v^4)$ the NRQCD Lagrangian describing a bound state between 
a quark of mass $m_1$ and an antiquark of mass $m_2$ is \cite{manohar}
\begin{eqnarray}
& &\!\!\!\!\!\!\!\!\!\!\!\!\!\!\!\! L = Q_1^\dagger\!\left(\!iD_0 + c^{(1)}_2 {{\bf D}^2\over 2 m_1} + 
c^{(1)}_4 {{\bf D}^4\over 8 m_1^3} + c^{(1)}_F g { {\bf \sigma}\cdot {\bf B} \over 2 m_1} 
+ c^{(1)}_D g { {\bf D}\!\cdot\!{\bf E} - {\bf E}\!\cdot\!{\bf D} \over 8 m_1^2} \right. 
\label{nrqcd}\\
& & \!\!\!\!\!\!\!\!\!\!\!\!\!\!\!\!
  \left. + i c^{(1)}_S g {{\bf \sigma} \!\!\cdot \!\!({\bf D}\!\times\!{\bf E} - {\bf E}\!\times\!{\bf D})
\over 8 m_1^2} \!\right)\!Q_1 + \hbox{\, antiquark terms}\, (1 \leftrightarrow 2)
+ {d_1\over m_1 m_2} Q_1^\dagger Q_2 Q_2^\dagger Q_1 
\nonumber\\
& & \!\!\!\!\!\!\!\!\!\!\!\!\!\!\!\!   + {d_2\over m_1 m_2}
 Q_1^\dagger {\bf \sigma} Q_2 Q_2^\dagger {\bf \sigma} Q_1
+ {d_3\over m_1 m_2} Q_1^\dagger T^a Q_2 Q_2^\dagger T^a Q_1 
+ {d_4\over m_1 m_2} Q_1^\dagger T^a {\bf \sigma} Q_2 Q_2^\dagger T^a {\bf \sigma} Q_1, 
\nonumber
\end{eqnarray}
where $Q_j$ are the heavy quark fields. Reparameterization invariance\cite{manohar}  
fixes $c_2=c_4=1$. 
 The coefficients $c^{(j)}_F$, $c^{(j)}_D$, ... are evaluated at the matching scale $\mu$ 
for a particle of mass $m_j$. They encode the ultraviolet regime of QCD order by order in $\alpha_{\rm s}$. 
The explicit expressions and a numerical discussion can be found in \cite{vairo}. 
The power counting rules  for the operators of Eq. (\ref{nrqcd}) are $Q \sim (mv)^{3/2}$, 
${\bf D} \sim mv$, $gA_0 \sim m v^2$, $g{\bf A} \sim m v^3$, $gE\sim m^2 v^3$ and 
$gB \sim m^2 v^4$. Four quark operators which are apparently of order $v^3$ are 
actually suppressed by additional powers in $\alpha_{\rm s}$ in the matching coefficients and the 
octet contributions by an additional power in $v^2$ on singlet states.  Therefore 
in the following we will neglect these contributions with the exception of a term which 
mixes under RG transformation with the chromomagnetic operator contribution to 
the spin-spin potential \cite{chen}. We will call the corresponding matching coefficient $d$. 
%%%%%%%%%%%%%%%%%%%%%%%%%%%%%%%%%%%%%%%%%%
\begin{figure}[htb]
\begin{centering}
\vskip -0.5truecm
\makebox[2truecm]{\phantom b}
\epsfxsize=6.1truecm
\epsffile{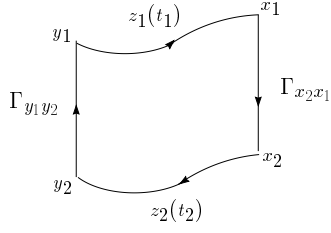}
\vskip -4.0truecm
\caption[x]{Distorted Wilson loop.}
\label{figuno}
\end{centering}
\end{figure}
\vspace{0.3cm}
%%%%%%%%%%%%%%%%%%%%%%%%%%%%%%%%%%%%%%%%%%%
\item{\it Step 2}.  Write down the quark-antiquark gauge-invariant Green function using a path integral 
 representation. The 4-point gauge invariant Green function $G$  
associated with the Lagrangian (\ref{nrqcd}) is defined as \cite{wilson,eichten} 
$$
G(x_1,y_1,x_2,y_2) =  \langle 0 \vert  Q_2^\dagger(x_2)  U(x_2,x_1) Q_1(x_1)  Q_1^\dagger(y_1)  U(y_1,y_2) 
Q_2(y_2) \vert 0\rangle, 
$$
where $U(x_2,x_1) \equiv \displaystyle\exp\left\{ - ig \int_0^1 ds \, (x_2-x_1)^\mu A_\mu(x_1 + s (x_2-x_1)) \right\}$ 
is a Schwi\-nger line added to ensure gauge invariance. After integrating out the heavy quark fields,  
$G$ can be expressed as a quantum-mechanical path integral over the quark trajectories:
\begin{eqnarray} 
& & \!\!\!\!\!\!\!\!\!\!\!\!\!\!\!\!\!
G(x_1,y_1,x_2,y_2) = \int_{y_1}^{x_1} \!\!\!{\cal D}z_1 {\cal D}p_1\int_{y_2}^{x_2} 
\!\!\!{\cal D}z_2 {\cal D}p_2 \exp \left\{i\int_{-T/2}^{T/2} \!\!dt \sum_{j=1}^2 {\bf p}_j\cdot {\bf z}_j -m_j -  {p_j^2\over 2 m_j} \right. 
\nonumber \\
& & \!\!\!\!\!\!\!
\quad  \left. +  {p_j^4\over 8 m_j^3} \right\}\, 
{1 \over N_c} \left\langle {\rm Tr \, P \, T_s} \exp\left\{ -ig \oint_\Gamma dz^\mu A_\mu(z) 
+i\int_{-T/2}^{T/2} \!\!dz_{0j} \, c^{(j)}_F  g {{\bf\sigma}\cdot{\bf B}\over 2 m_j}  \right.\right. 
\nonumber \\
& & \!\!\!\!\!\!\!
\quad \left.
+ c^{(j)}_D g { {\bf D}\!\cdot\!{\bf E} - {\bf E}\!\cdot\!{\bf D} \over 8 m_j^2} 
+ i c^{(j)}_S g {{\bf \sigma} \!\!\cdot \!\!({\bf D}\!\times\!{\bf E} 
- {\bf E}\!\times\!{\bf D}) \over 8 m_j^2} \right\}
\label{green}\\
& & \!\!\!\!\!\!\!
\quad \left.
\times \exp\left\{ {i\over m_1m_2} \int_{-T/2}^{T/2} \!\!dt \, g^2 d\, T^{a(1)} {\bf \sigma}^{(1)} 
T^{a(2)} {\bf \sigma}^{(2)}\delta^3({\bf z}_1 - {\bf z}_2) \right\} \right\rangle \equiv
\nonumber \\
& & \!\!\!\!\!\!\!\!\!\!\!\!\!\!\!\!
 \int_{y_1}^{x_1} \!\!\!{\cal D}z_1 {\cal D}p_1\int_{y_2}^{x_2} \!\!\!{\cal D}z_2 {\cal D}p_2
\exp\left\{i\int_{-T/2}^{T/2} \!\!\!\!\! dt \sum_{j=1}^2 {\bf p}_j\cdot {\bf z}_j -m_j - 
{p_j^2\over 2 m_j}  + {p_j^4\over 8 m_j^3} -i \int_{-T/2}^{T/2} \!\!\!\!\! dt \, U\right\}, 
\nonumber
\end{eqnarray}
where the bracket means the Yang--Mills average over the gauge fields, $\Gamma$ is the Wilson loop 
made up by the quark trajectories $z_1$ and $z_2$  and the endpoints Schwinger strings and 
$y_2^0 = y_1^0 \equiv -T/2$, $x_2^0 = x_1^0 \equiv T/2$, see Fig. \ref{figuno}. 
\item{\it Step 3}. Extract the form of the heavy quark interaction from Eq. (\ref{green}).
Assuming that the limit exists, we define the heavy 
quark-antiquark potential $V$ as $\displaystyle \lim_{T\to\infty} \int_{-T/2}^{T/2} \!\!dt \,U /T$. 
Expanding in $v$ we get 
\begin{eqnarray}
& & V(r) = \lim_{T \to \infty} { i \log \langle W(\Gamma) \rangle  \over T} +
\label{pot}\\
& & \!\!\!\!\!\!\!\!\!\!\!\!  \left( {{\bf S}^{(1)}\cdot{\bf L}^{(1)}\over m_1^2} +  
{{\bf S}^{(2)}\cdot{\bf L}^{(2)}\over m_2^2} \right)\!
{2 c^+_F V_1^\prime(r) + c^+_S V_0^\prime(r) \over 2r} 
+ { {\bf S}^{(1)}\cdot{\bf L}^{(2)}  + 
{\bf S}^{(2)}\cdot{\bf L}^{(1)}  \over m_1 m_2} {c^+_F V_2^\prime(r) \over r}
\nonumber\\
& & \!\!\!\!\!\!\!\!\!\!\!\!\!\!\!\!\!\!\!\!
+ \left( {{\bf S}^{(1)}\cdot{\bf L}^{(1)}\over m_1^2} - 
{{\bf S}^{(2)}\cdot{\bf L}^{(2)}\over m_2^2} \right) \!
{2 c^-_F V_1^\prime(r) + c^-_S V_0^\prime(r) \over 2r} 
+ { {\bf S}^{(1)}\cdot{\bf L}^{(2)}  - 
{\bf S}^{(2)}\cdot{\bf L}^{(1)}  \over m_1 m_2} {c^-_F V_2^\prime(r) \over r}
\nonumber\\
& & \!\!\!\!\!\!\!\!\!\!\!\!
+{1\over 8}\left( {c_D^{(1)} \over m_1^2} 
+ {c_D^{(2)} \over m_2^2} \right) (\Delta V_0(r) + \Delta V_{\rm a}^E(r)) 
+{1\over 8}\left( {c_F^{(1)} \over m_1^2} 
+ {c_F^{(2)} \over m_2^2} \right) \Delta V_{\rm a}^B(r)
+{c_F^{(1)}c_F^{(2)}\over m_1 m_2} \times
\nonumber\\
& & \!\!\!\!\!\!\!\!\!\!\!\!
\left( 
{{\bf S}^{(1)}\!\cdot\!{\bf r} {\bf S}^{(2)}\!\cdot\!{\bf r} \over r^2} - 
{{\bf S}^{(1)}\!\cdot \!{\bf S}^{(2)} \over 3} \right) \! V_3(r) 
+ {{\bf S}^{(1)}\!\cdot\!{\bf S}^{(2)} \over 3 m_1 m_2}
\left( c_F^{(1)} c_F^{(2)} V_4(r) -48 \pi \alpha_{\rm s} C_F \, d \, \delta^3(r)\right).
\nonumber
\end{eqnarray}
where 
$
W(\Gamma) \equiv  {\rm P \,} \displaystyle\exp\left\{ -ig \oint_\Gamma dz^\mu A_\mu(z) \right\}
$
 is the averaged value of the deformed Wilson loop, see Fig. \ref{figuno}.
  The expansion of it around the static Wilson loop $W(\Gamma_0)$  
($\Gamma_0$ is a $r\times T$ rectangle) gives the static potential 
$V_0 = \displaystyle\lim_{T \to \infty}  i \log \langle W(\Gamma_0) \rangle / T$ 
plus velocity (non-spin) dependent terms \cite{nora,vel} which are 
controlled by four scale-independent potentials $V_i(r)\,\, i=b,c,d,e$.  ${\bf S}^{(j)}$ and ${\bf L}^{(j)}$ 
are the spin and orbital angular momentum operators of the particle $j$. 
The matching coefficients are defined as $2 c^{\pm}_{F,S} \equiv c^{(1)}_{F,S} \pm c^{(2)}_{F,S}$.
The spin-dependent potentials agree with the ones obtained in refs. \cite{eichten} with the 
exception of the matching coefficients that were introduced in \cite{chen}.
\end{itemize}
All the potentials $V_1-V_4\, V_a-V_d$ are obtained as  functions of the  average value of 
the distorted Wilson loop $\langle W(\Gamma)\rangle$ and  insertions of one or two 
field strengths  in this 
average, $\langle F_{\mu\nu}  W(\Gamma)\rangle$  or $\langle F_{\mu\nu} F_{\rho\sigma} W(\Gamma)\rangle $.
Via deformation of the quark or (antiquark) trajectory, such v.e.v. of field strength insertion can 
be calculated via functional derivatives of the Wilson loop.  We conclude that to obtain the complete 
quark-antiquark order $O(v^4)$ interaction  (quenched) no other assumptions are needed than the behavior of
$\langle W(\Gamma)\rangle$: given $\langle W(\Gamma)\rangle$ everything is analytically calculable.
On the other side, expanding the average of the distorted Wilson loop $\Gamma$ in terms of the 
static Wilson loop $\Gamma_0$
  we get expressions for the potentials suitable for lattice evaluations\cite{potlat,balipot}.
  In this way,  we can compare 
unambiguously the predictions for the heavy quark interaction obtained  in  various QCD vacuum models
 and the lattice measures as well as the phenomenological data.

\section{ QCD vacuum models}

Models of the QCD vacuum are needed in order to describe 
  the nonperturbative behavior of the Wilson loop 
average. 
To this aim one wants to exploit all the available lattice information on the mechanism of confinement and 
 all  the measurements of the  Wilson loop.
Let us consider the v.e.v. of the Wilson loop. It pays to expand this average 
in terms of field strength expectation values, by using the non-Abelian Stokes theorem \cite{svm}
\begin{eqnarray}
& & \!\!\!\!\!\!\!\!\!  \langle W(\Gamma)\rangle\equiv \langle \exp
\{ig  \oint_\Gamma dz^\mu A_\mu(z)\} \rangle
=\langle P \exp ig \int_{S(\Gamma)} dS^{\mu\nu}(1) U(0,1)
   F_{\mu\nu}(1) U(1,0)\rangle = 
\nonumber \\
& & \!\!\!\!\!\!\!\!\!\!\!  \exp \big \{\sum_{n=0}^\infty {(ig)^n \over n!} 
\int_{S(\Gamma)} dS^{\mu_1\nu_1}(1) \cdots  dS^{\mu_n\nu_n}(n)  
\langle U(0,1)
 F_{\mu_1\nu_1}(1) \cdots
  F_{\mu_n\nu_n}(n)U(n,0) \rangle_{\rm cum} \big \}
\nonumber
\end{eqnarray}
where $i\equiv x_i$, $S(\Gamma)$ denotes a surface with contour $\Gamma$ and $\langle
 \cdots\rangle_{\rm cum}$ 
stands for the cumulant average. In principle, the v.e.v. of the Wilson loop is given 
by the sum of all the  cumulants. However, in a recent lattice  investigation
(\cite{balinoi} and refs. therein), 
evidence of the Gaussian dominance in the cumulant expansion of quasi-static Wilson loop 
averages was found. Therefore, it follows that 
\begin{equation}
\langle W(\Gamma) \rangle \simeq \exp \big \{ {-{1\over 2}} \int_{S(\Gamma)} dS^{\mu\nu}(0) \int_{S(\Gamma)} 
dS^{\rho\sigma}(x)
\langle g^2 U(0,x)F_{\mu\nu}(x)U(x,0)F_{\rho\sigma}(0) \rangle \big \}
\label{gauss}
\end{equation}
is a good approximation. This is  the basic assumption in the (Gaussian) stochastic vacuum model
 \cite{svm} and it was   phenomenologically confirmed by calculation in high energy scattering 
and quarkonia. Then, the heavy quark interaction is  determined by the two-point field 
strength correlator
\begin{eqnarray}
&& \!\!\!\!\!\!\!\!\! g^2 
\langle U(0,x) F_{\mu\nu}(x) U(x,0) F_{\lambda\rho} (0) \rangle =
  \ {g^2 \langle F^2(0)\rangle \over 24 N_c } 
\bigg\{ (g_{\mu\lambda}g_{\nu\rho} - 
g_{\mu\rho}g_{\nu\lambda})(D(x^2) + D_1(x^2)) 
\nonumber \\
&&  + (x_\mu x_\lambda g_{\nu\rho} - 
x_\mu x_\rho g_{\nu\lambda} 
+ x_\nu x_\rho g_{\mu\lambda} - x_\nu x_\lambda g_{\mu\rho})
{d\over dx^2}D_1(x^2) \bigg\} .
\label{due}
\end{eqnarray}
The Lorentz decomposition is general and the dynamics is contained in the form factors $D$ and $D_1$.
The function $D$ is responsible for the  area law and confinement (indeed in QED, due to the  Bianchi identity,  we
have $D=0$). For $D$ and $D_1$
the lattice calculations \cite{balinoi,digiaco}
 give an exponential (in Euclidean space)
 long-range decreasing behavior  $\simeq G_2 \exp\{-\vert x\vert /T_g\}$, 
where $G_2\equiv \langle \alpha_s F^2(0)\rangle/\pi $ is the gluon condensate and $T_g\simeq
0.15\div  0.2\,  {\rm fm}$
\footnote{Phenomenological calculation in high energy scattering indicates $T_g\equiv 0.3\div0.35 \, 
{\rm fm}$ }
 is the gluon correlation length (quenched). \par
In ref. \cite{abel}, the QCD two--point field strength correlator (\ref{due}) has been related to the dual field propagator of the effective Abelian Higgs model describing infrared QCD. In this way the Gaussian dominance
in the Wilson loop average is understood as following from the classical approximation
 in the dual theory. Moreover, it is possible to relate the QCD parameter $T_g$ and
$G_2$ to the dual parameters.
 In the London limit $T_g$ is identified with the dual gluon mass $M$,
without the London limit the relation is more involved but still $T_g$ is expressed in terms of the dual theory 
parameters. \par 
From the calculation of the heavy quark interaction in  various model of the  QCD vacuum (minimal area
 law model \cite{vel}, stochastic vacuum model \cite{svm,mod}, dual QCD \cite{dual,mod,abel}, Isgur and 
Paton model \cite{isgur})  we can state that:
\begin{itemize}
\item{ All these models give the same result for the nonperturbative heavy quark interaction not 
only in the long range regime but also in the transition region.}
\item{ Two nonperturbative parameters, that can be related to $T_g$ and $G_2$,  control 
the nonperturbative interaction.}
\item{ All these models predict the Eichten-Feinberg nonperturbative spin interaction 
(pure Thomas precession) in the limit $T_g/r \to 0$. In the transition region there is a 
subleading correction to the spin-interaction coming from the magnetic interaction.}
\item{All these models predict the nonperturbative velocity dependent corrections to be proportional to the 
flux tube angular momentum. This prediction is definitively different from the result obtained 
with the semi-relativistic reduction of Bethe-Salpeter kernels of the type  $1/Q^4$,
 ($Q$ being the momentum transfer)
with any Lorentz structure, see \cite{nora}.}
\end{itemize}
The conclusion is that we need two parameters $T_g$ and $G_2$ to describe the heavy quark dynamics and indeed 
they are necessary to control the structure of the flux tube. Had we only one parameter, like the string tension 
$\sigma$, we could  encode the information of a constant energy density in the flux tube. However,
the whole structure is important, and also the information about
 the width of the flux tube has to be considered\footnote{The flux tube distribution between 
a static quark-antiquark pair is measured on the lattice, see \cite{nora}}.
 In the limit of very large inter-quark distances 
 and in particular dynamical regimes, we can store the relevant information in one parameter, the string 
tension.  For instance, from Eqs. (\ref{gauss}) and (\ref{due}), 
 the confining part of the static potential is \cite{svm}
\begin{equation}
V_0(r) \simeq  G_2 \int_0^\infty d\tau
\int_0^r d \lambda \, (r-\lambda) D(\tau^2 +\lambda^2)
\label{svpot} 
\end{equation}
and the string tension  $\sigma$ emerges
 as an integral on the $D$ function  $\sigma \simeq G_2$ $  \int_0^\infty d\tau $ $ 
\int_0^\infty  d\lambda$ $  D(\tau^2 +\lambda^2) $ in the limit ${T_g/ r}\to 0$.

\section{Heavy-light systems}

\begin{figure}[htb]
%\begin{centering}
\vskip -1truecm
\makebox[1truecm]{\phantom b}
\put(0,-50){\epsfxsize=6.5cm \epsfbox{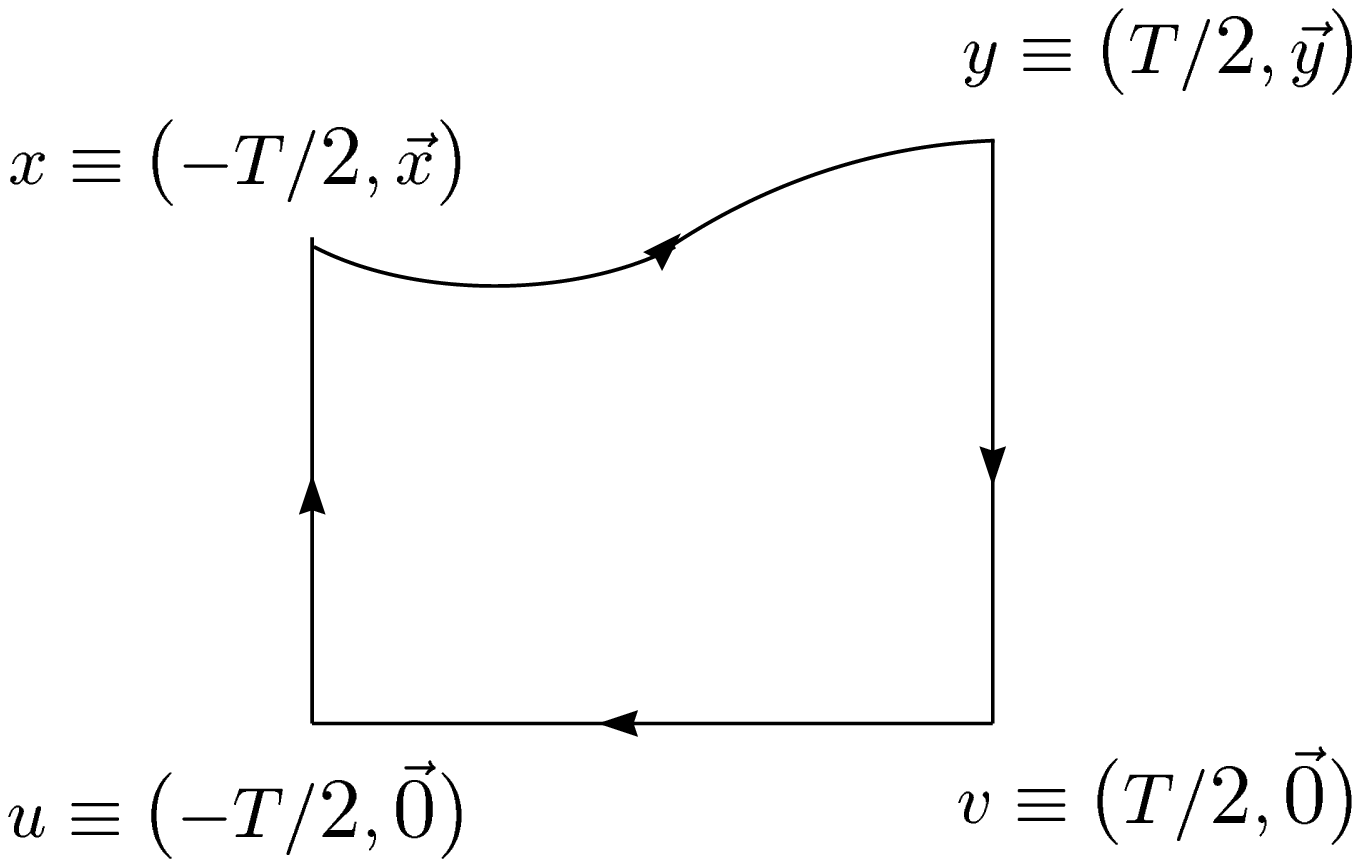}}
\put(220,70){\epsfxsize=4.5cm \epsfbox{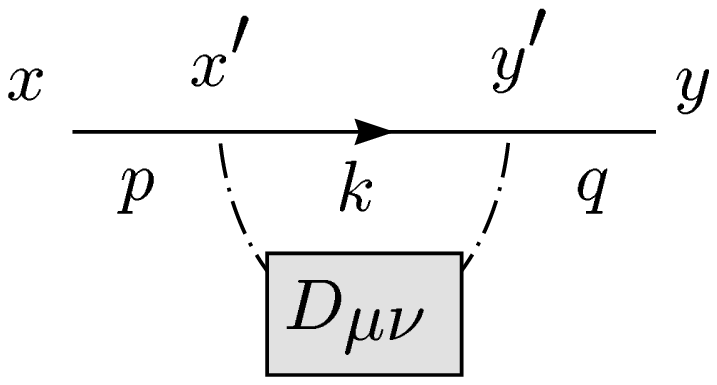}}
\vskip -2truecm
\caption[x]{The Wilson loop in the static limit of the heavy quark and the interaction kernel $K$.}
\label{figundici}
%\end{centering}
\end{figure}

We study the heavy-light bound state  system in the non-recoil limit.
We start from the gauge-inva\-riant quark-antiquark
 Green function in the Fey\-nman-Schwi\-nger re\-pre\-sen\-tation
\cite{dirac}:
\begin{eqnarray}
&~&\!\!\!\!\!\!\!\!\!\!\!\!\! G(x,u,y,v) =
{1\over 4} \Bigg\langle {\rm Tr}\,{\rm P}\, 
(i\,{D\!\!\!\!/}_{y}^{\,(1)}+m_1)\, 
\int_{0}^\infty dT_1\int_{x}^{y}{\cal D}z_1 \,
e^{\displaystyle - i\,\int_{0}^{T_1}dt_1 {m^2+\dot z_1^2 \over 2}   }
\nonumber\\
&~&\times
\int_{0}^\infty dT_2\int_{v}^{u}{\cal D}z_2 \,
e^{\displaystyle - i\,\int_{0}^{T_2}dt_2 {m^2+\dot z_2^2 \over 2}   }
e^{\displaystyle ig \oint_\Gamma dz^\mu A_\mu(z)}
\label{feyschwi}\\
&~&\times 
e^{\displaystyle i\,\int_{0}^{T_1}dt_1 \, {g\over 4}\sigma_{\mu\nu}^{(1)}
F^{\mu\nu}(z_1)}
e^{\displaystyle i\,\int_{0}^{T_2}dt_2 \, {g\over 4}\sigma_{\mu\nu}^{(2)}
F^{\mu\nu}(z_2)} 
(-i\,\buildrel{\leftarrow}\over{D\!\!\!\!/}_{v}^{\,(2)} + m_2) \Bigg\rangle . 
\nonumber
\end{eqnarray} 
Again the dynamics is contained in the Wilson loop, that now looks like
 Fig. \ref{figundici}.
%%%%%%%%%%%%%%%%%%%%%%%%%%%%%%%%%%%%%%%%%%%
%%%%%%%%%%%%%%%%%%%%%%%%%%%%%%%%%%%%%%%%%%%
 We can 
exploit the symmetry of the situation, taking the  modified coordinate gauge
$
 A_\mu(x_0,{\bf 0}) = 0$, $ x^jA_j(x_0,{\bf x}) = 0$.
Notice that this gauge choice is possible due to the gauge-invariance of 
the formalism. Within this gauge is possible to express the gauge field in 
terms of the field strength tensor $ A_\mu(x) =\int_0^1 d\alpha \alpha^{n(\mu)} x^k F_{k\mu}(x_0, \alpha {\bf x}) $
where $n(0)=0, n(i)=1$. Then, the only non-vanishing contribution  to the Wilson loop is
\begin{equation}
W(\Gamma) =    
{\rm  Tr \,} {\rm  P\,} \exp \left\{ ig \int_x^y dz^\mu A_\mu (z) \right\}.
\label{wilh}
\end{equation}
At this point, at variance from the heavy quark case, we have to make a  model dependent assumption, i. e. 
we consider  still valid the  dominance of the bilocal correlator.
Indeed, this should be  a property of the vacuum.
Then, we have 
\begin{eqnarray}
\langle W(\Gamma) \rangle & \simeq & 
\exp \left\{ - {g^2\over 2} \int_x^y dx^{\prime\mu} \int_x^y dy^{\prime\nu}
D_{\mu\nu}(x^\prime,y^\prime) \right\},\label{proph}\\
D_{\mu\nu}(x,y) &\equiv& 
x^ky^l\int_0^1 d\alpha \, \alpha^{n(\mu)} \int_0^1 d\beta\, \beta^{n(\nu)}
\langle
 F_{k\mu}(x^0,\alpha{\bf x})F_{l\nu}(y^0,
\beta{\bf y})\rangle \nonumber
\end{eqnarray}
and inserting Eq. (\ref{proph}) in (\ref{feyschwi}) and expanding the exponential we obtain 
the following expression for the propagator $S_D$ of the light  quark in the static heavy quark field:
\begin{equation}
S_D = S_0 + S_0 \, K \, S_0 + S_0 \, K \, S_0 \, K \, S_0 + \cdots ,  
\label{sdexp}
\end{equation} 
$S_0$ being the free fermion propagator. Taking into account only the first 
planar graph (since we are interested only in contributions proportional to 
the gluon condensate), we have $K(y^\prime,x^\prime) = \gamma^\nu 
S_0(y^\prime,x^\prime) \gamma^\mu D_{\mu\nu}(x^\prime,y^\prime)$. 
A graphical representation of $K$ is given in Fig. \ref{figundici}.
Eq. (\ref{sdexp}) can be written in closed form  
as $S_D = S_0 + S_0 K S_D$ (or in terms of the wave-function, 
$({p\!\!\!/} -m - iK)\psi = 0$; $m\equiv m_1$). Therefore, $K$ can 
be interpreted as the interaction kernel of the Dirac equation associated 
with the motion of a quark in the field generated by an 
infinitely heavy antiquark. \par
 Notice that \cite{dirac}: 1) $K$ {\it is not a translational invariant quantity}. 
The coordinate gauge breaks 
explicitly this symmetry in the propagator. Physically this is due to the presence of the heavy quark.
2) The kernel depends on $D_{\mu\nu}$ which in turns is given in terms of the two-point correlator (\ref{due}).
Then, the heavy-light dynamics is controlled by the same two parameters controlling the heavy-heavy dynamics, $T_g$ and $G_2$.
3) The problem has many relevant scales:  the light  mass $m$,
 the correlation length $T_g\sim \Lambda_{QCD}$,
 the characteristic energy and momentum of the bound state. We have
 different dynamical regimes in dependence on the relative 
values of these scales. Let us study the various situations. 
In the following we consider only the nonperturbative dynamics \cite{dirac}.

\begin{itemize}
\item{ {\it Potential Case:  $m>{1/ T_g} > p_0-m, {\bf p}, {\bf p}-{\bf q}$.} 
We neglect the negative energy states  and expand the kernel $K$ 
 in $m$. We  obtain: 
$$
\!\!\!\!\!\!\!\!\!\!\!\! V(r)\sim G_2 \big \{ \int_{-\infty}^{+\infty}
\!\!\!\!\!\! d\tau \int_0^r d \lambda(r-\lambda) 
D(\tau^2+\lambda^2) + {{\bf \sigma}\cdot {\bf L} \over 4 m^2 r} 
\int_{-\infty}^{+\infty}\!\!\!\!\!\!\! d\tau \int_0^r d \lambda 
\left( {2 \lambda\over r} - 1\right) D(\tau^2+\lambda^2)\big\}
$$
which coincides in the limit of large $r$ with the Eichten-Feinberg 
potential \cite{eichten} $ V(r) = \sigma r 
- {{\bf \sigma}\cdot {\bf L} \over 4 m^2} \,\, {\sigma \over r},$ 
with   $\sigma$  defined  as in Sec. 3.
We emphasize that the Lo\-rentz structure  which gives origin to the negative 
sign in front of the spin-orbit potential (hence to the pure  Thomas precession term)  is in our 
case {\it not simply a scalar}  ($K\simeq \sigma \, r$). }
\item{{\it Sum Rules case: $({1/T_g}<p_0-m$, ${1/T_g}<m)$.}
We get  the well-known  Shifman-Vainshtein-Zakharov result for the heavy quark condensate 
$$
\langle \bar{Q} Q \rangle  = 
-\int {d^4p \over (2\pi)^4}\int {d^4q \over (2\pi)^4} 
{\rm Tr} \left\{ S_0(q)K(q,p)S_0(p) \right\} = -{1\over 12} 
{\langle \alpha F^2(0) \rangle \over \pi m}.
$$ }
\item{{\it  $D_s$ and $B_s$ case: ${1/T_g} >m$.} 
 For $D_s$ and $B_s$ one can probably  still assume that the propagator inside 
the kernel is free and solve the equation to get the spectrum.}
\item{{\it $D$ and $B$ case: $  m\ll {1/T_g}$.} 
We observe that in the zero mass limit the kernel $K$
gives a chirally symmetric interaction (while a purely scalar interaction 
breaks chiral symmetry at any mass scale). This means on one side that our 
interaction keeps the main feature of QCD i.e. in the zero mass limit 
chiral symmetry is broken only spontaneously. On the other side 
this seems to suggest that for very light quarks the quark propagator 
 should be taken 
from the chiral broken solution, i.e. the nonlinear equation 
\cite{simonov,cond} 
\begin{equation}
S_D=S_0+S_0 K(S_D)S_D
\label{schwind}
\end{equation}
has to be solved with Schwinger--Dyson like techniques.}
\end{itemize}
In conclusion our approach to the heavy-light systems  contains 
the sum rules and the potential results still  allowing for a chiral symmetric interaction.
In the next section we discuss the interplay between confinement and chiral symmetry breaking that emerges 
in this picture.

\section{Confinement and chiral symmetry breaking}

We address the problem of solving Eq. (\ref{schwind}) \cite{cond}.
 We will see how   chiral symmetry breaking 
emerges in a heavy-light bound state and how it leads to a non--chiral invariant 
confining interaction. In place of  (\ref{proph})
 we consider the simplified interaction 
\begin{equation}
g^2 D_{\mu\nu}(x,y) \simeq - i {\delta_{\mu 0}\delta_{\nu 0} \over 24}
T_g \langle g^2 E^2(0)\rangle \,  {\bf x} \cdot {\bf y} \,\delta(x_0-y_0)
\label{rough}
\end{equation}
where we have considered the leading contribution 
as  given by the electric fields only depending on time 
  and  we have  approximated the  exponential fall off in time (with correlation length $T_g$) 
with an instantaneous delta-type interaction \footnote{This can be done since for light quarks the energy scale $T_g^{-1}$ is expected to be bigger
than the other scales of the problem.}.\par
We find convenient to work in the Hamiltonian
 approach. 
 The effective Hamiltonian  corresponding to Eqs. (\ref{schwind}) and (\ref{rough}) 
 is:
\begin{eqnarray}
& & \!\!\!\!\!\!\!\!\!\!\!\!\!\!\!
 H  = \int d^3x \, q^\dagger({\bf x}) (-i {\bf \alpha} \cdot {\bf \nabla} +m\beta) q({\bf x})  
- {1\over 2} \int d^3x \int d^3y \, V^3_0 r^2 \,  q^\dagger({\bf x}) T^a  q({\bf x}) 
q^\dagger({\bf y}) T^a  q({\bf y}) 
\nonumber\\
& & + 2 \int d^3x \int d^3y \, V^3_0 R^2  \, q^\dagger({\bf x}) T^a  q({\bf x}) 
q^\dagger({\bf y}) T^a  q({\bf y}), 
\label{heff2}
\end{eqnarray}
with $
V_0^3 \equiv -{T_g }\langle g^2 E^2(0)\rangle/96  =  {T_g } \pi^2 G_2/96 $ 
and  ${\bf R} = {\bf x}/2 + {\bf y}/2$ and ${\bf r} = {\bf x} - {\bf y}$.  
From Eq. (\ref{heff2}) it is  clear the role played by the approximation (\ref{rough}). 
It  allows  to disentangle trivially in the  effective Hamiltonian the self interacting 
part (function of $r$) from the external source interacting term (function of $R$). 
With a ``realistic'' lattice  parameterization of the non-local gluon 
condensate  these terms might be mixed up in a very complicate way. 
\par
By means of the  Bogoliubov--Valatin variational method we  select 
the chiral broken vacuum. The quark fields are expanded  on a trial basis of spinors:  
$
q({\bf x})$ $  = \sum_s \int {d^3k \over (2\pi)^3}  e^{i {\bf k}\cdot {\bf x}}
\left [ u_s({\bf k}) b_s({\bf k}) + v_s({\bf k})d^\dagger_s(-{\bf k})  \right ],
$
with  trial spinors $u_s$ and $v_s$,
$
u_s({\bf k})$ $= {1\over \sqrt{2}}$ $  [ \sqrt{1 + \sin \phi(k)}$
  $+ \sqrt{1 - \sin \phi(k)} {\bf \alpha}\cdot \hat{\bf k} ]u^0_s$,
$v_s({\bf k})$ $= {1\over \sqrt{2}}$ $  [ \sqrt{1 + \sin \phi(k)}$
 $ - \sqrt{1 - \sin \phi(k)} {\bf \alpha}\cdot \hat{\bf k} ]v^0_s$,
where $u^0_s$ and $v^0_s$ are the usual rest-frame spinors on the chiral invariant vacuum. 
In the limiting case  $\phi=0$ the trial spinors reduce to the massless free one, while for 
$\phi = \pi/2$ they reduce to infinitely  massive  sources.\par
 Expanding the Hamiltonian (\ref{heff2}) on the trial basis  we get $
H = {\cal E} + H^r_2 + H^R_2+ H_4$, where 
\begin{eqnarray}
{\cal E} &=& {\cal V} \left\{ 3 \int {d^3k \over (2\pi)^3} {\rm Tr} \left[ 
({\bf \alpha} \cdot {\bf k} + m \beta) \Lambda_-({\bf k}) \right] \right. 
\nonumber \\
& & \left . -  2 V_0^3 \int {d^3k \over (2\pi)^3} \int {d^3k' \over (2\pi)^3} 
{\rm Tr} \left[  \Lambda_-({\bf k}) \Lambda_+({\bf k'}) \right] 
\int d^3r e^{-i({\bf k} - {\bf k'}) \cdot {\bf r} } r^2 \right\}, 
\label{ee}\\
H^r_2 &=& \int d^3x : q^\dagger({\bf x}) \left( -i {\bf \alpha}\cdot {\bf \nabla} + m \beta \right) 
q({\bf x}):
\nonumber\\
& & -{2\over 3} V_0^3 \int d^3R \int d^3r  \int {d^3k \over (2\pi)^3}  r^2  e^{i{\bf k} \cdot {\bf r} } 
:q^\dagger({\bf x}) \left[ \Lambda_+({\bf k}) - \Lambda_-({\bf k}) \right] q({\bf y}):  
\nonumber \\
H^R_2 &=& {8\over 3} V_0^3 \int d^3R \int {d^3k \over (2\pi)^3}  R^2  e^{i{\bf k} \cdot {\bf r} } 
:q^\dagger({\bf x}) \left[ \Lambda_+({\bf k}) - \Lambda_-({\bf k}) \right] q({\bf y}):  
\nonumber\\
H_4 &=& - {1\over 2} V_0^3 \int d^3R \int d^3r  (r^2 - 4 R^2)
:q^\dagger({\bf x}) T^a q({\bf x}) q^\dagger({\bf y}) T^a q({\bf y}): 
\nonumber
\end{eqnarray}
where $: ~~:$ is the normal ordering operator and $\cal V$ is the volume of the space. 
$\cal E$ is the vacuum energy, $H_2^r$  is the light quark kinetic energy on the physical vacuum, 
$H_2^R$ is the binding interaction
 (as far as the bare $q\bar{Q}$ mass is concerned we do not need to evaluate 
$H_4$ matrix elements).
\begin{itemize}
\item{ {\it The gap equation} 
$\delta {\cal E}(\phi) = 0.$
Explicitly  it reads 
\begin{equation}
m \cos\phi(k) - k\sin\phi(k) + {2\over 3} V_0^3 \left(\phi(k)^{\prime\prime} + {2\over k}\phi(k)^\prime 
+ {2\over k^2} \cos\phi(k) \sin\phi(k) \right) = 0.
\label{gap2}
\end{equation}
The light quark condensate can be calculated and gives:$
\langle 0 \vert \bar{q} q \vert 0 \rangle$ 
$ = -{3\over \pi^2}$ $ \int_0^\infty dk $ $ k^2  \sin\phi(k). 
$
On the solution of Eq. (\ref{gap2}) we get for $m=0$   
\begin{equation}
\langle 0 \vert \bar{q} q \vert 0 \rangle = -{1\over 24} \, T_g \, G_2 \times 0.3722 
\label{qbarq}
\end{equation}
{\it The result (\ref{qbarq}) is appealing. It establishes a connection between the gluon condensate 
and the light quark condensate. The connection is possible since the non-local gluon condensate  
has introduced into the game a finite correlation length $T_g$.} }
\item{ {\it The bound state equation}.
Taking the matrix element of $H_2^R$  between  
a one-particle state of momentum ${\bf p}$ and a one-particle state of momentum ${\bf q}$, we have  
\begin{eqnarray}
& &H_2^R({\bf p},{\bf q})_{ss'} \equiv 
\langle 0 \vert b_s({\bf p}) H_2^R  b^\dagger_{s'}({\bf q}) \vert 0 \rangle = 
\nonumber\\
& & \!\!\!\!\! {8\over 3} V_0^3 
u^\dagger_s({\bf p}) \left\{  \beta \sin\left[\phi\!\left({{\bf p} + {\bf q} \over 2}\right)\right]
+ {{\bf\alpha}\cdot\hat{\bf p} +{\bf\alpha}\cdot\hat{\bf q} \over 2} 
\cos\left[\phi\!\left( {{\bf p} + {\bf q} \over 2}\right) \right] \right\}u_{s'}({\bf q})
\nonumber\\
& & \qquad\qquad\qquad \times \left[ -\Delta \, (2\pi)^3\delta^3({\bf p} - {\bf q}) \right] .
\nonumber
\end{eqnarray}
As expected the binding interaction would be chiral invariant ($\sim {\bf\alpha} \cdot\hat{\bf p} 
+ {\bf\alpha}\cdot\hat{\bf q}$) for a massless particle  on the perturbative vacuum ($\phi = 0$). 
While for a infinitely massive particle ($\phi = \pi/2$) chiral invariance would be  maximally broken. 
In our case the solution of the gap equation (\ref{gap2}) gives rise to a binding interaction 
which contains two pieces. One is chiral invariant and the other, proportional to $\beta$, breaks explicitly 
chiral invariance. The existence of such a term is suggested by the spin-orbit structure of the heavy quarkonium  
potential whose relativistic origin may be traced back to a scalar confining Bethe--Salpeter kernel. 
In a Hamiltonian language this would just correspond to an interaction proportional to $\beta$. 
On the contrary, here  we obtain 
 an interaction not only 
proportional to $\beta$. It manifests, also under the strong simplifying assumption (\ref{rough}), 
a more complicate structure which interpolates between a chiral invariant 
vector interaction and a scalar interaction.

Summing up the contributions coming from the pieces $H_2^r$ and $H_2^R$ of the Hamiltonian, the bound state 
equation on the physical vacuum reads
$$
\!\!\!\!\!\!\!\!\!\!\!
\left\{ E(p) + {8\over 3} \,V_0^3\left(  {1\over 2 p^2} [1-\sin\phi({\bf p})]^2 
 + {2\over p^2}[1-\sin\phi({\bf p})]
\, {\bf S}\cdot{\bf L}  - \Delta \right) \right\}  \Phi({\bf p}) = \bar{\Lambda} \,  \Phi({\bf p}),
$$
where we have introduced the spin operator ${\bf S} = {\bf \sigma}/2$ and the orbital
 angular momentum operator 
${\bf L} = {\bf r}\times{\bf p}$. The eigenvalues $\bar{\Lambda}$ of the equation are the energy levels of 
the bound state in the non-recoil limit, i.e. the difference
 between the mass  of the considered 
heavy-light meson and the mass  of the corresponding heavy quark.}
\end{itemize}

\section{Conclusions}
We have described the heavy quark and the heavy-light quark bound state  system using the same 
gauge-invariant approach and in terms of the two nonperturbative parameters $T_g$ and $G_2$.
 Chiral symmetry breaking and a chiral non-invariant binding interaction emerge quite 
naturally in our approach and a link is established between chiral symmetry breaking properties 
and confining interaction. In particular with Eq. (\ref{qbarq}) we establish a relation 
between the order parameter of chiral symmetry (the quark condensate $\langle 0 \vert \bar{q} q \vert 0\rangle$) 
and that one which in our framework describes confinement (the gluon correlation length $T_g$). 
The actual calculations were performed under the rough approximation (\ref{rough}). 
This is unrealistic since it gives in the heavy quark limit a confining potential 
which is not linear. Moreover all magnetic contributions were not considered. Nevertheless  we expect 
that the main features presented   will still hold  using a
realistic parameterization of the bilocal gluon condensate.
\par
\noindent{\bf Acknowledgements}
 It is a pleasure for N. B.  to thank the organizers  for the perfect organization of the Conference,
  the invitation and  the warm hospitality.  N. B. acknowledges the support of the 
European Community, Marie Curie fellowship, TMR Contract n. ERBFMBICT961714.

\end{document}